\documentclass[reprint,aip,jcp,amssymb]{revtex4-2}

\usepackage[tbtags]{amsmath}
\usepackage{graphicx}
\usepackage{bm}

\newcommand{\fig}[1]{Fig.\,\ref{fig:#1}}

\newcommand{\pr}{\partial}

\makeatletter

\renewcommand\frontmatter@abstractwidth{\dimexpr\textwidth\relax}
\makeatother

\usepackage[hidelinks]{hyperref} 

\begin{document}

\title{
Extrapolating molecular dynamics simulations to zero time step and across thermodynamic space}

\author{Kush Coshic}
\affiliation{
Department of Theoretical Biophysics, Max Planck Institute of Biophysics, 60438 Frankfurt am Main, Germany
}

\author{Gerhard Hummer}
\email{gerhard.hummer@biophys.mpg.de}
\affiliation{
Department of Theoretical Biophysics, Max Planck Institute of Biophysics, 60438 Frankfurt am Main, Germany
}
\affiliation{
Institute of Biophysics, Goethe University Frankfurt, 60438 Frankfurt am Main, Germany}

\begin{abstract}
The integration time step is a critical determinant of performance in molecular dynamics simulations, governing the trade-off between speed and fidelity. 
Although 2 fs remains the standard in atomistic biomolecular simulations, the push for performance has popularized a 4 fs time step with hydrogen mass repartitioning, often combined with multiple time stepping or mass rescaling.
However, it is often unclear whether a chosen protocol is overly aggressive, as the apparent numerical stability of a trajectory can mask underlying thermodynamic inaccuracies.
Increasing the time step will exacerbate systematic discretization errors, inherent to all numerical integration algorithms.
In the widely used Verlet family of integrators, these errors manifest as $\mathcal{O}(\Delta t^2)$ deviations in thermodynamic observables such as potential energy and volume, and for common Langevin splitting schemes, even temperature.
We demonstrate that these deviations follow a simple, linear thermodynamic model, allowing for their rigorous removal by extrapolation to the zero time step limit. In turn, the time-step dependence provides us with estimates of the system heat capacity, compressibility, and thermal expansion coefficient. This framework allows us to construct consistent probability distributions of energy and volume across thermodynamic states, effectively recovering Boltzmann-consistent statistics at a target condition independent of time step.
These considerations are particularly important for enhanced sampling methods such as replica exchange and umbrella sampling, which rely on rigorous Boltzmann sampling and require accurate energies and temperatures for valid replica exchange probabilities and statistical reweighting.
\end{abstract}
\maketitle

\section{Introduction}

Despite tremendous hardware and software developments, atomistic molecular dynamics (MD) simulations of biomolecular systems remain largely restricted to microsecond timescales. 
The fundamental bottleneck is the integration time step, which is constrained by the fastest molecular motions in the system. For time steps that are too large, numerical integration becomes unstable due to hard particle collisions or rapid bond vibrations that generate large forces and cause the trajectory to diverge. 
For conventional biomolecular simulations, a 2 fs time step provides a sufficient safety buffer against numerical instability, ensuring that integration failures remain vanishingly unlikely over accessible timescales.\cite{Oberhofer:PRE:2007,ROSA2024}

A variety of workarounds have been developed to stabilize the numerical integration of the equations of motion with large time steps.
For instance, the highest-frequency motions, typically bond stretching involving hydrogen atom, are often frozen using holonomic constraint algorithms.\cite{Ryckaert:JCP:1977,ANDE1983,MIYA1992,HESS1997}
By removing these high-frequency modes from the dynamics, constraint methods enabled a doubling of the time step from 1 fs to the current community standard of 2 fs. 
Hydrogen mass repartitioning (HMR)\cite{HOPK2015} redistributes mass from heavy atoms to bonded hydrogens, thereby slowing bond vibrations and enabling larger integration time steps. 
In a complementary approach, rescaling the mass of solvent atoms enhances conformational sampling by reducing solvent viscosity.\cite{walser1999validity,walser2001viscosity,LIN2010}
Recent work\cite{ROSA2024} combines HMR with mass scaling to leverage the benefits of both approaches.
HMR has become widely adopted in biomolecular simulations, enabling stable 4 fs time step integration.\cite{BALU2019}

Another commonly used strategy to enhance performance is multiple time-stepping (MTS),\cite{Streett1978,GRUB1991,TUCK1992} whereby the total force is decomposed into rapidly varying (short-range) and slowly varying (long-range) components. 
In principle, the slowly varying forces can be evaluated less frequently and applied as impulses over longer intervals, thereby reducing the computational cost. 
In practice, however, the maximum impulse interval is limited by resonance artifacts, where coupling to high-frequency modes can induce severe instability.\cite{MA2003,ABRE2021}

While increasing the time step offers clear gains in simulation throughput, concerns persist regarding both the physical fidelity and numerical accuracy of such approaches. On the one hand, artificially increasing hydrogen mass, or even freezing hydrogen motion via constraints can suppress conformational transitions that depend on hydrogen dynamics, potentially hindering the exploration of configurational space and hence sampling efficiency.
On the other hand, all finite-time step integrators introduce systematic discretization errors that grow with time step size~($\Delta t$) and can bias thermodynamic observables even when trajectories remain numerically stable.\cite{LEIM2015}
When choosing a time step for an MD simulations, we thus trade-off sampling speed against accuracy. In practice, the emphasis has been on sampling speed, with biomolecular simulations typically being performed with time steps close to the stability of the numerical time integration.\cite{Lechner:JCP:2006,Oberhofer:PRE:2007,Fabian:JCP:2022,ROSA2024} However, the resulting boost in sampling efficiency comes with the risk of systematic errors in the calculated quantities.

How large are the errors from integrating MD trajectories with finite time steps? Generally speaking, any numerical integration scheme involving discretization does not exactly preserve the invariants of the continuous system but instead propagates a modified system.
For the widely used Verlet-family integrators, which are time-reversible and second-order accurate, the modified system differs by terms of second order in the time step, $\mathcal{O}(\Delta t^2)$. 
Provided the integrator is symplectic, the numerical solution conserves a modified ``shadow" Hamiltonian that stays close to the true energy manifold rather than randomly drifting away.\cite{LEIM2015}
Seminal work has characterized these errors for a variety of integrators, splitting schemes, and thermostat couplings.\cite{LEIM2013}
While the Verlet integrator propagates an $\mathcal{O}(\Delta t^2)$ error in both positions and velocities, how the error translates to thermodynamic quantities is more subtle, depending on splitting schemes and thermostat couplings.
For instance, the Brünger-Brooks-Karplus (BBK) scheme,\cite{BRUN1984} the default for Langevin dynamics in the widely used MD engines NAMD and AMBER, produces a systematic $\mathcal{O}(\Delta t^2)$ temperature bias, causing the simulated temperature to deviate from the thermostat setpoint and complicating efforts to sample the intended canonical ensemble.\cite{PAST1988}
The deviation is amplified at larger time steps, and is especially concerning at 4 fs.
While the BAOAB scheme mitigates the $\mathcal{O}(\Delta t^2)$ error in temperature,\cite{LEIM2013B} an $\mathcal{O}(\Delta t^2)$ error persists in other quantities such as energies and volume.
Hybrid Monte Carlo sampling \cite{Duane:1987} offers an alternative to sample from the desired ensemble and is thus used for applications desiring high accuracy \cite{Menzl:2016} or for sampling from a modified potential.\cite{siggel:trimem:2022}
Textbooks and reviews on molecular simulation discuss  issues of sampling accuracy in detail, providing guidelines for stable and accurate integration.\cite{LEIM2015,allen2017}

Despite the theoretical foundation, the practical implications of discretization errors are often underappreciated by practitioners, particularly in the context of Langevin dynamics. 
For stochastic thermostats that rescale velocities or kinetic energies, such as the Bussi–Donadio–Parrinello thermostat,\cite{BUSS2007} temperature control is maintained explicitly, mitigating the discretization-induced temperature drift at least to $\mathcal{O}(\Delta t^2)$. 
Langevin dynamics, however, couples each degree of freedom to an implicit heat bath via friction and stochastic forces, and the discretized equations of motion can produce a system temperature that deviates systematically from the thermostat setpoint in a time step-dependent manner. 
Ensuring that simulations are performed in the correct thermodynamic ensemble is essential for obtaining reliable thermodynamic and kinetic properties, yet ad hoc tuning of the thermostat setpoint to compensate for discretization bias is both cumbersome and potentially problematic. 

Here we derive a thermodynamic framework that relates the integration time step to measured observables. We use this formalism to extrapolate MD simulation to the zero time step limit. As an added benefit, we extract thermodynamic information from the time-step dependence, in particular on the heat capacity, compressibility, and thermal expansion coefficient.
We then show that this model provides the necessary correction terms to recover the correct Boltzmann distribution from trajectories biased by large integration time steps.
Such corrections are critical for enhanced sampling techniques like replica exchange MD\cite{Okamoto:CPL:1999:remd}  and umbrella sampling,\cite{Torrie:74} which rely on precise knowledge of the system temperature and energies; neglecting the integration-induced drifts may lead to incorrect Boltzmann weighting and, as shown for inadequate thermostatting,\cite{Rosta:JCTC:2009} flawed thermodynamic conclusions.

\section{Methods}

\subsection{Thermodynamic model}

We write the Gibbs free energy, $G(p,T)$ as an expansion to second order in pressure ($p$) and temperature ($T$), about a reference state ($p_0$,$T_0$),
\begin{equation}
\begin{split}
G(p, T) = {} & G_0
+ \bigg(\frac{\pr G}{\pr p}\bigg)_{T_0}(p-p_0)
+ \bigg(\frac{\pr G}{\pr T}\bigg)_{p_0}(T-T_0) \\
& + \frac{1}{2}\bigg(\frac{\pr ^2 G}{\pr p^2}\bigg)_{T_0}(p-p_0)^2 
+ \frac{1}{2}\bigg(\frac{\pr ^2 G}{\pr T^2}\bigg)_{p_0}(T-T_0)^2 \\
& + \bigg(\frac{\pr ^2 G}{\pr p {} \pr T}\bigg)_{p_0,T_0}(p-p_0)(T-T_0) 
\end{split}
\label{eq:gibbs0}
\end{equation}
where we ignored terms of order three and higher. Here, $C_p$ is the heat capacity at constant pressure, $\kappa _T$ is the isothermal compressibility, and $\alpha$ is the thermal expansion coefficient, defined as:
\begin{align}
    C_p &=\ T \bigg(\frac{\pr S}{\pr T}\bigg)_{p} 
\label{eq:Cp}\\
    \kappa_T &= - \frac{1}{V} \bigg(\frac{\pr V}{\pr p}\bigg)_{T} 
\label{eq:kappa}\\
    \alpha &= \frac{1}{V} \bigg(\frac{\pr V}{\pr T}\bigg)_{p} 
\label{eq:alpha}
\end{align}
where $S$ is the system entropy. 
Using the Maxwell relations, we rewrite $G(p,T)$ as:
\begin{equation}
\begin{split}
    G(p,T) = {} & G_0 + (p-p_0)V_0 - S_0(T-T_0) - \frac{1}{2}\kappa_TV_0(p-p_0)^2\\
    & -\frac{C_p}{T_0}(T-T_0)^2 + \alpha V_0 (p-p_0)(T-T_0)
\end{split}
\label{eq:gibbs}
\end{equation}
From Eq.~\eqref{eq:gibbs}, we get the entropy $S$ as:
\begin{align*}
  S(p,T) = - \left( \frac{\pr G}{\pr T} \right)_{P_0} = S_0 - \alpha V_0 (P-P_0) - \frac{C_p}{T_0}(T-T_0)
\end{align*}
and using the relation $U = G + TS - pV$, we get an expression for the internal energy $U$ as:
\begin{equation}
\begin{split}
    U(p,T) = {} & U_0 + \frac{1}{2}\kappa_TV_0(p^2-p_0^2) - \alpha V_0 (pT - p_0T_0)\\
    & + \frac{C_p}{2T_0} (T-T_0)(T+T_0)
\end{split}
\label{eq:U_0}
\end{equation}
Linearizing Eq.~\eqref{eq:U_0} in $p$, $T$ gives:
\begin{equation}
\begin{split}
    U(p,T) = {} & U_0 + \big(C_p - \alpha V_0 p_0\big) (T-T_0) \\[1.13ex]
    & + \big( \kappa_T V_0 p_0 - \alpha V_0 T_0\big) (p-p_0)
\end{split}
\label{eq:U_1}
\end{equation}
Equation~\eqref{eq:U_1} is the linearized approximation for the internal energy. The analogous linear expression for the volume $V$ is:
\begin{align}
    V(p,T) = \left( \frac{\pr G}{\pr p} \right)_{T_0} = V_0 - \kappa_TV_0 (p-p_0) + \alpha V_0(T-T_0)
    \label{eq:V_0}
\end{align}
Following a similar procedure, all other thermodynamic quantities can be written to linear order in $p$ and $T$. 
An important quantity is the configurational enthalpy ($H_\mathrm{conf}$) of the system,\cite{allen2017} defined as:
\begin{equation}
    H_\mathrm{conf} = U_\mathrm{pot} + p V
\label{eq:H}
\end{equation}
where $U_\mathrm{pot}$ is the potential energy of the system. The configurational enthalpy refers to the potential energy contribution to the enthalpy ($H$), i.e., excluding the kinetic energy contribution. 
Thermostats maintain temperature by artificially controlling the kinetic energy by means of velocity rescaling or stochastic forces. At constant temperature in the classical (non-quantum) limit, the resulting kinetic energy contribution to the enthalpy is simply a constant term ($\frac{1}{2}fk_{B}T$ with $N$ the number of degrees of freedom; $f=3N$ for $N$ atoms).
We will use $H_\mathrm{conf}$ for correcting trajectories across different combinations of \{$\Delta t, T_\mathrm{set}, p_\mathrm{set}$\}, to validate the accuracy and robustness of the thermodynamic model.

\subsection{Time step dependence in the thermodynamic model}

Within this thermodynamic framework, we assume that we sample from a potential surface that is slightly perturbed in manner that depends on the simulation time step.  Verlet-like integrators introduce a discretization error in the propagated phase-space coordinates  $\left(\mathbf{r}(t),\mathbf{p}(t)\right)$ for positions ($\mathbf{r}$) and momenta ($\mathbf{p}$) that scales to leading order as $\mathcal{O}(\Delta t^{2})$. Generally, ensemble averages of observables that are smooth functions of these coordinates—such as the potential energy $U_\mathrm{pot}(\mathbf{r})$, kinetic energy $K(\mathbf{p})$, or total energy $U(\mathbf{r},\mathbf{p})$—will inherit this quadratic dependence on $\Delta t$, though higher-order terms may become relevant for large $\Delta t$, depending on the molecular interactions and composition of the system. 
Here, we posit that in the standard simulation regime, the sampled thermodynamic state is dominated by this leading-order bias, resulting in a systematic and predictable $\mathcal{O}(\Delta t^{2})$ shift in the effective stationary distribution that is distinct from random statistical fluctuations.
This systematic shift in the ensemble-averaged thermodynamic state serves as the explicit basis for our thermodynamic correction and extrapolation model.

For the family of Verlet integrators with a Langevin like thermostat coupling, the physical temperature of the simulated system ($T$) will be slightly perturbed from the temperature set in the thermostat ($T_\mathrm{set}$). 
While the BBK scheme will theoretically produce an ($\mathcal{O}(\gamma \Delta t)$) deviation, where $\gamma$ is the collision frequency,\cite{PAST1988} this term is negligible for commonly used values for the collision frequency (0.5-5~ps$^{-1}$).
Empirically, we find that the leading order correction is $\mathcal{O}(\Delta t^{2})$, and hence write the system temperature as:
\begin{equation}
    T = T_\mathrm{set} + a_T \Delta t^2
\label{eq:T}
\end{equation}
By incorporating a leading-order $\mathcal{O}(\Delta t^{2})$ discretization error into the thermodynamic expansions for $U(p,T)$ and $V(p,T)$, Eqs.~\eqref{eq:U_1} and \eqref{eq:V_0}, we obtain:
\begin{equation}
\begin{split}
    U(p,T) = {} & U_0 + a_U \Delta t^2 + \big(C_p - \alpha V_0 p_0\big) (T-T_0) \\[1.13ex]
    & + \big( \kappa_T V_0 p_0 - \alpha V_0 T_0\big) (p-p_0) 
\end{split}
\label{eq:U}
\end{equation}
\begin{equation}
    V = V_0 + a_V \Delta t^2 + \alpha V_0(T-T_0) - \kappa_TV_0 (p-p_0) 
\label{eq:V}
\end{equation}
Here, $a_T$, $a_U$, and $a_V$ are coupling constants.
We find empirically that the simulated system pressure ($p$) remains very close to the barostat setpoint ($p_{\mathrm{set}}$), such that any systematic time step dependence in $p$ can be ignored.
Hence, we make the assumption that $p \approx~p_\mathrm{set}$ in Eqs.~\eqref{eq:U} and \eqref{eq:V}. 
Equations~\eqref{eq:T}, \eqref{eq:U}, and \eqref{eq:V} constitute our final thermodynamic model, in which we incorporate systematic finite time step corrections and approximate the thermodynamic observables to linear order in the variables $p$ and $T$.

The time step dependent drift in the thermodynamic variables ($V, T$) implies that we are effectively sampling a modified potential surface at a modified thermodynamic state. 
Provided the integrator is symplectic, the discrete deterministic dynamics conserves a shadow Hamiltonian ($\tilde{\mathcal{H}}$), that approximates the true Hamiltonian ($\mathcal{H}$) with an error of order $\mathcal{O}(\Delta t^2)$:
\begin{equation*}
    \tilde{\mathcal{H}}(\mathbf{p},\mathbf{r}) = \mathcal{H} (\mathbf{p},\mathbf{r}) + \mathcal{O}(\Delta t^2)
\end{equation*}
The system thus evolves on a stable effective energy surface that remains close to the physical one, in effect preventing long-term energy drift, assuming there are no other errors.
In the limit $\Delta t \rightarrow 0$, we recover thermodynamic observables corresponding to the true Hamiltonian, thereby eliminating the systematic artifacts introduced by the finite integration time step.
Historically, one has had to resort to Monte Carlo methods to sample strictly from a specified thermodynamic ensemble for a given system. Here, we find that by performing a small number of simulations and fitting the measured observables to a thermodynamic model, one can instead reliably extrapolate to the zero time step limit across a range of thermodynamic conditions.

\subsection{Fitting the model}

The thermostat temperature setpoint ($T_\mathrm{set}$) and the simulation time step ($\Delta t$) are the independent variables in the model. 
The resulting temperatures $T (\Delta t,T_\mathrm{set})$, $U(\Delta t, T_\mathrm{set}, p)$, and volumes $V(\Delta t, T_\mathrm{set}, p)$ are measured from the simulation and used to globally fit the 8 fitting parameters in the model, i.e., \{$U_0$, $\kappa_T$, $V_0$, $C_p$, $\alpha$, $a_U$, $a_V$, $a_T$ \}. 
Among these, $\alpha$, $\kappa_T$ and $C_p$ are standard physical observables with reasonable a priori estimates for many systems. 
Here, we set the reference pressure ($p_0$) to 1 bar and reference temperature ($T_0$) to 310 K. 
The fitted quantities $U_0(T_0,p_0)$ and $V_0(T_0,p_0)$ are the internal energy and volume of the system at the reference state ($T_0,p_0$), respectively, in the limit of $\Delta t \rightarrow 0$.
The coefficients \{$a_T$, $a_U$, $a_V$\} depend on both the system and the numerical discretization scheme.

Model parameters were determined via a weighted global non-linear least-squares optimization using \texttt{scipy.optimize.least\_squares}.\cite{VIRT2020} The analytic expressions for temperature, total energy, and volume--Eqs. \eqref{eq:T}, \eqref{eq:U}, and \eqref{eq:V}, respectively--were fitted simultaneously to the MD simulation data, ensuring parameter consistency across all observables. 
To account for the disparate units and magnitudes, residuals for each observable were normalized by the standard deviation of that observable prior to minimization. 
Parameter uncertainties are reported as asymptotic standard errors, given by the square roots of the diagonal elements of the scaled covariance matrix $s^2(J^\top J)^{-1}$, where $J$ is the Jacobian of the normalized residuals with respect to the model parameters evaluated at the optimal solution, and $s^2 = \chi^2/(N-p)$ is the residual variance for $N$ observations and $p = 8$ free parameters.
Below, we demonstrate that global fits to a series of independent MD simulations reliably yield fitting coefficients that accurately describe the data across a wide range of pressures and temperatures. This robust performance underscores the distinct advantage of formulating the model as linear in $p$ and $T$, thereby enabling reliable extrapolation to the zero time step limit. If a wide thermodynamic range is covered by the simulations, beyond the range of the linear approximation, thermodynamic models of higher order in $p$ and $T$ can be used instead in a global fit.

\subsection{Correcting energy, enthalpy, and volume distributions to recover the correct thermodynamic ensemble in the limit of zero time step}
Enhanced sampling methods such as replica exchange and umbrella sampling explicitly assume that configurations are sampled from Boltzmann distributions for the given potential energy,
\begin{equation}
    p(\mathbf{p},\mathbf{r},V)\propto \exp(-\beta \mathcal{H}(\mathbf{p},\mathbf{r})-\beta pV)
\end{equation}
where $\beta=1/(k_B T)$. However, as discussed, the time step error in time integration induces errors of order $\mathcal{O}(\Delta t^2)$. In the best case, one would have Boltzmann sampling from the shadow Hamiltonian and weight the configurations accordingly, 
\begin{equation}
    p(\mathbf{p},\mathbf{r},V)\propto \exp(-\tilde{\beta} \tilde{\mathcal{H}}(\mathbf{p,r})-\tilde{\beta} p\tilde{V})
\end{equation}
but this is usually not practical, lacking simple explicit expressions for the shadow Hamiltonian, $\tilde{\mathcal{H}}$ and the possibly altered reciprocal temperature, $\tilde{\beta}$.
Here, we instead use the thermodynamic relations, Eqs. \eqref{eq:T}, \eqref{eq:U}, and \eqref{eq:V}, to correct for the time step error and to estimate the Boltzmann distribution at zero time step and the thermodynamic conditions of interest.
In practice, this is done by translating (“shifting”) the sampled distributions from a finite time step source simulation to the desired target state ($\Delta t_{\text{ref}}=0, T_{\text{ref}}=310~\mathrm{K}, p_{\text{ref}}=1~\mathrm{bar}$).

Given a source trajectory sampled at ($\Delta t_{\text{set}}, T_{\text{set}}, p_{\text{set}}$) with instantaneous total energies $U_{\text{sim}}(t)$ and volumes $V_{\text{sim}}(t)$, and with time averaged temperature $T_{\text{sim}}$ and pressure $p_{\text{sim}}$, we first use the thermodynamic model to compute the expected shifts to a target state ($\Delta t_\text{ref},T_\text{ref},p_\text{ref})$,
\begin{equation}
\begin{aligned}
\Delta U = {} & a_U (\Delta t_{\text{ref}}^2 - \Delta t_{\text{sim}}^2)
+ \big(C_p - \alpha V_0 p_0\big) (T_{\text{ref}} - T_{\text{sim}}) \\
& + \big(\kappa_T V_0 p_0 - \alpha V_0 T_0\big) (p_{\text{ref}}-p_{\text{sim}}) \\[1.1ex]
\Delta V = {} & a_V (\Delta t_{\text{ref}}^2 - \Delta t_{\text{sim}}^2)
 + \alpha V_0(T_{\text{ref}}-T_{\text{sim}})\\
&- \kappa_T V_0 (p_{\text{ref}}-p_{\text{sim}}) 
\end{aligned}
\label{eq:deltaUV}
\end{equation}
We then correct the total energy and volume by adding these shifts,
\begin{equation}
\begin{split}
    U_{\text{corrected}}(t) = U_{\text{sim}}(t) + \Delta U \\
    V_{\text{corrected}}(t) = V_{\text{sim}}(t) + \Delta V
\end{split}
\end{equation}
We illustrate the correction procedure using the configurational enthalpy, $H_{\text{conf}}(\mathbf{r}, V)$, a quantity not explicitly represented in the model's thermodynamic relations, Eqs. \eqref{eq:T}, \eqref{eq:U}, and \eqref{eq:V}. 
However, standard thermodynamic relations link $H_{\text{conf}}$ to quantities appearing in the model ($U, V, T$), so corrections derived in terms of these variables can be transferred directly to $H_{\text{conf}}$ and, by the same logic, to other derived quantities of interest.
For the configurational enthalpy $H_{\text{conf}}$, we subtract the kinetic energy from the total energy,
\begin{equation}
    H_{\text{conf}}(t)=U_{\text{sim}}(t)-\frac{f}{2}k_B T_\text{sim}+p_\mathrm{ref}V(t)
\end{equation}
For simple systems like rigid water, the classical degrees of freedom $f$ are known analytically ($f=6$ per water molecule). For more complex systems with overlapping holonomic constraints, analytically evaluating $f$ can become cumbersome, although, in principle, it can be obtained by graph-theoretic rigidity/constraint counting methods for biomolecules.~\cite{JACO2001} In MD engines, it is usually assumed that every added constraint removes one degree of freedom, which may, however, not be the case. 
Here, we accordingly determine an effective $f$ from an arbitrarily chosen reference simulation assuming equipartition, calculated from the ratio of the reported average kinetic energy $K$ to the average temperature $T_\text{sim}$, i.e., 
$f = 2\langle K \rangle / k_B T_\text{sim}$. 
Since $f$ is an intrinsic property of the system, we use the calculated value to analytically correct for kinetic energy contributions across different thermodynamic states.
Then we correct the total energy and volume using Eq.~\eqref{eq:deltaUV},
\begin{equation}
\begin{split}
    H_{\text{conf, corrected}}(t) = {} & H_{\text{conf}}(t)+\Delta U -\frac{f}{2}k_B (T_{\text{ref}} - T_{\text{sim}}) \\
    & + p_\text{ref}\Delta V
\end{split}
\label{eq:H_corrected}
\end{equation}
The term $\frac{f}{2}k_B (T_{\text{ref}} - T_{\text{sim}})$ subtracts the shift in kinetic energy, which is needed to remove the kinetic contribution inherently included in the total energy correction $\Delta U$. 
For $\Delta t_\text{ref}=0$, $H_{\mathrm{conf, corrected}}(t)$ gives a Boltzmann-distributed sample of the zero time step Hamiltonian $\mathcal{H}$ at $(T_{\mathrm{ref}}, p_{\mathrm{ref}})$. Our thermodynamic mapping effectively corrects for the $\mathcal{O}(\Delta t^{2})$ macroscopic energy difference between the true Hamiltonian $\mathcal{H}$ and the shadow Hamiltonian $\tilde{\mathcal{H}}$ natively sampled by the finite time step integrator, under the assumptions that errors are of second order and that linear thermodynamics can be used to correct for deviations from the target state.
We will show that this analytical mapping is robust and consistent; with the coefficients established, mapping configurational enthalpy from disparate source simulations yields zero time step distributions in exact correspondence, effectively eliminating the discretization bias. 

The mapping scheme is generalizable: provided an observable is explicitly captured by our linear thermodynamic model, such as the potential energy $U$ or the system volume $V$, its probability distribution can be similarly translated to a target state without relying on statistical reweighting. Non-thermodynamic physical observables can be included in the thermodynamic model. To linear order in $T$ and $p$, and quadratic order in $\Delta t$, the resulting correction for, say, the mean end-to-end distance $R$ of a peptide is
\begin{equation}
    \Delta R = a_R (\Delta t_{\text{ref}}^2 - \Delta t_{\text{sim}}^2)
+ R_T (T_{\text{ref}} - T_{\text{sim}}) + R_p (p_{\text{ref}}-p_{\text{sim}})
\end{equation}
The additional coefficients $a_R$, $R_T$, and $R_p$ can be determined by fitting the time step, temperature, and pressure dependence of $R$. We expect the fit parameters $R_T$ and $R_p$ to be the derivatives of the mean $R$ with respect to $T$ and $p$ in the thermodynamic ensemble,
\begin{equation}
    R_T = \left(\frac{\partial \langle R\rangle_{p,T}}{\partial T}\right)_p; \quad R_p = \left(\frac{\partial \langle R\rangle_{p,T}}{\partial p}\right)_T
\end{equation}
in complete analogy to the thermodynamic model above.

\subsection{Simulation methods}

We fitted the thermodynamic model on MD simulations of (i) pure water and (ii) a protein system comprising a ubiquitin (PDB ID: 1UBQ) placed in aqueous solution. 
All simulations were performed using NAMD 3.0.2,\cite{PHIL2020} the CHARMM-modified parameters for TIP3P water model (mTIP3P)\cite{MACK1998} and CHARMM36m for protein.\cite{HUAN2017}
SETTLE\cite{MIYA1992} and RATTLE\cite{ANDE1983} algorithms were applied to constrain covalent bonds to hydrogen in water and in non-water molecules, respectively. 
The temperature was maintained using a Langevin thermostat with a damping constant of 0.5~ps$^{-1}$.
Constant pressure (NPT) simulations employed a Nos\'e-Hoover Langevin piston barostat\cite{MART1994} with a period and decay of 200 and 100~fs, respectively.
The Langevin equations of motion were integrated using the method of Brünger, Brooks, and Karplus, as the default implementation in NAMD.\cite{BRUN1984,PHIL2005}
Energy minimization was carried out using the conjugate gradients method.\cite{PAYN1992}
Two different systems were simulated: (i) pure water in a triclinic box comprising of 5,184 water molecules and box lattice parameters $a = b = 46$ Å, $c = 86.3$ Å, and angles $\alpha = \beta = 90^\circ$, $\gamma = 120^\circ$, (ii) a ubiquitin molecule in water comprising in total 32,260 atoms in a cubic box of 69.5 {\AA} width. The total mass of the protein system ($\approx$~194898.198~amu) was used to make the observables intensive, with volume reported in units of ({\AA}$^3$ amu$^{-1}$) and energies in units of (J/g).
All systems were generated using VMD\cite{HUMP1996} and analyses performed with MDanalysis.\cite{AGRA2011}
Parameter fitting for the thermodynamic model was conducted using a series of simulations that varied in set temperature ($T_\mathrm{set}$), pressure ($P_\mathrm{set}\approx~p$), and integration time step ($\Delta t$). 
The full set of simulation parameters and averaged observables are listed in Table S1, S2 and S3 of the Supporting Information.

\section{Results}

\subsection{Systematic drift in thermodynamic properties}

\begin{figure}[!htbp]
\centering
\includegraphics[width=0.48\textwidth]{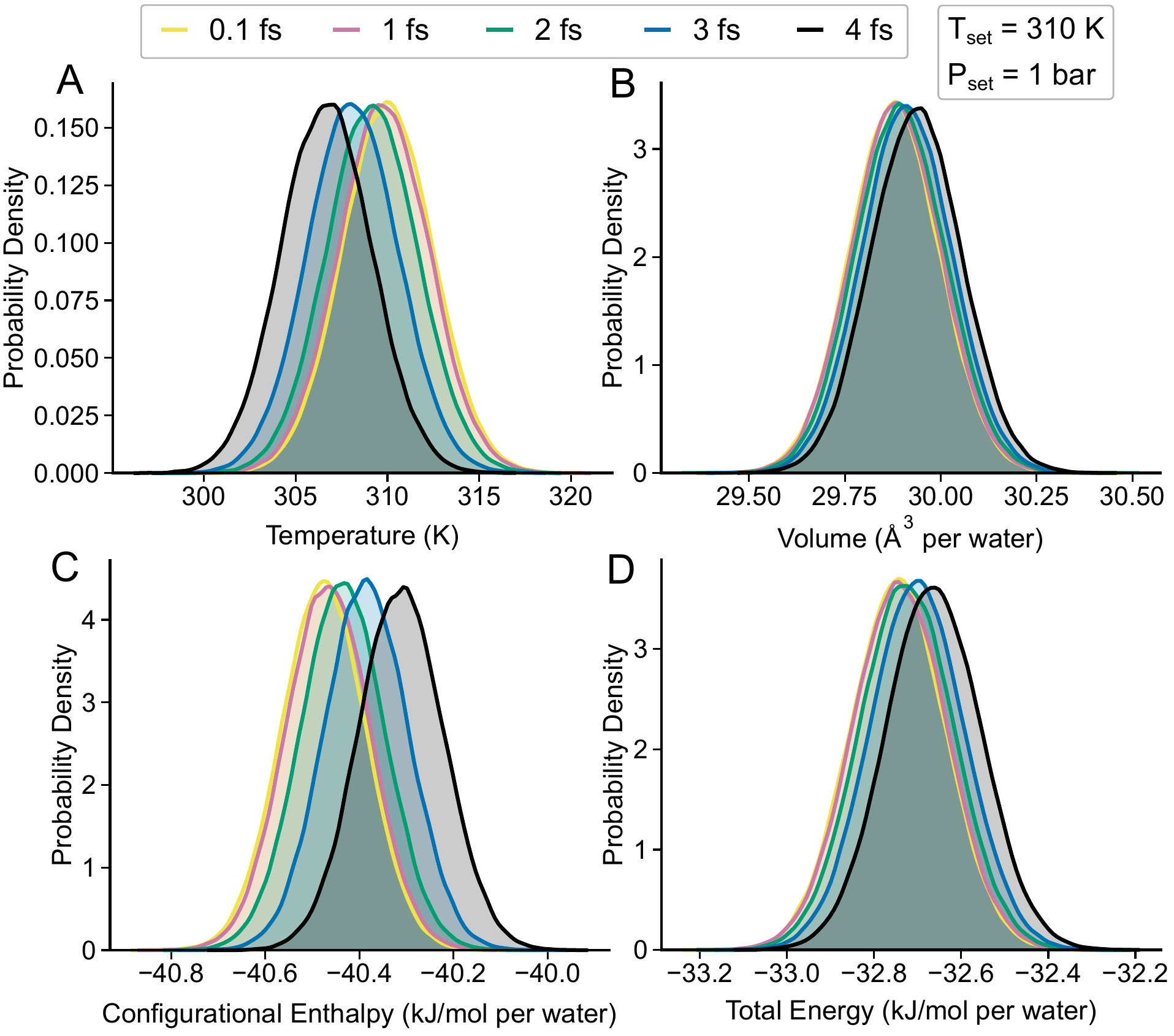}
\vspace{-4mm}
\caption{
\textbf{Thermodynamic properties depend on the time step.}
Probability density distributions of (A) temperature, (B) system volume, (C) configurational enthalpy, and (D) total energy obtained from simulations of 5,184 TIP3P water molecules, with varying integration time steps ($\mathrm{\Delta t}$) as per legend. Results in (B-D) are per water molecule. Simulations were performed for 1~$\mathrm{\mu}$s at a set temperature, $T_\mathrm{set}=310$ K and pressure, $P_\mathrm{set}=1$ bar.
For smoothing, the normalized distributions are presented as Gaussian kernel density estimates.
}
\label{fig:fig1}
\end{figure}

We first simulate a simple system of 5,184 water molecules in a triclinic box and demonstrate that changing the integration time step induces a systematic drift in all thermodynamic properties~(\fig{fig1}). 
For generality, the energies and volume are made intensive, evaluating them per water molecule.
The system temperature consistently falls below the thermostat set point in a time-step-dependent manner. This deviation becomes increasingly pronounced at larger time steps, as illustrated by the probability densities for the instantaneous temperature (\fig{fig1}A). The shift follows the quadratic dependence on $\Delta t$, given by Eq.~\eqref{eq:T}: it is already significant at 2~fs (0.8~K) and grows to 1.87~K at 3~fs and 3.33~K at 4~fs.
While such a quadratic trend is well known for the BBK integrator, the specific magnitude of the drift is determined by the system's highest-frequency motions. Consequently, pure water simulations—dominated by rapid O–H bond stretching—serve as a practical upper bound for these errors; aqueous solutions of, say, proteins should exhibit similar deviations.

Thermodynamic observables other than $T$ also exhibit systematic dependences on the simulation time step. We find a noticeable increase in the system volume with increasing $\Delta t$~(\fig{fig1}B), albeit with a relatively smaller drift. However, despite water's near-incompressibility under ambient conditions, this time step-dependent volumetric expansion is non-negligible and has to be explicitly incorporated into the fitting model, Eq.~\eqref{eq:V}.
Energetic properties are similarly affected, though in a more complex manner. 
As $\Delta t$~increases, interatomic collisions become effectively harder, causing the potential energy to rise. This is reflected in the pronounced increase of the configurational enthalpy with $\Delta t^2$~(\fig{fig1}C).
However, the simultaneous decrease in temperature reduces the kinetic energy contribution. Consequently, while the total energy still increases with time step, the magnitude of this shift is attenuated relative to the configurational enthalpy alone~(\fig{fig1}D).

\subsection{Fitting the thermodynamic model to MD simulations}

\begin{figure}[!htbp]
\centering
\includegraphics[width=0.48\textwidth]{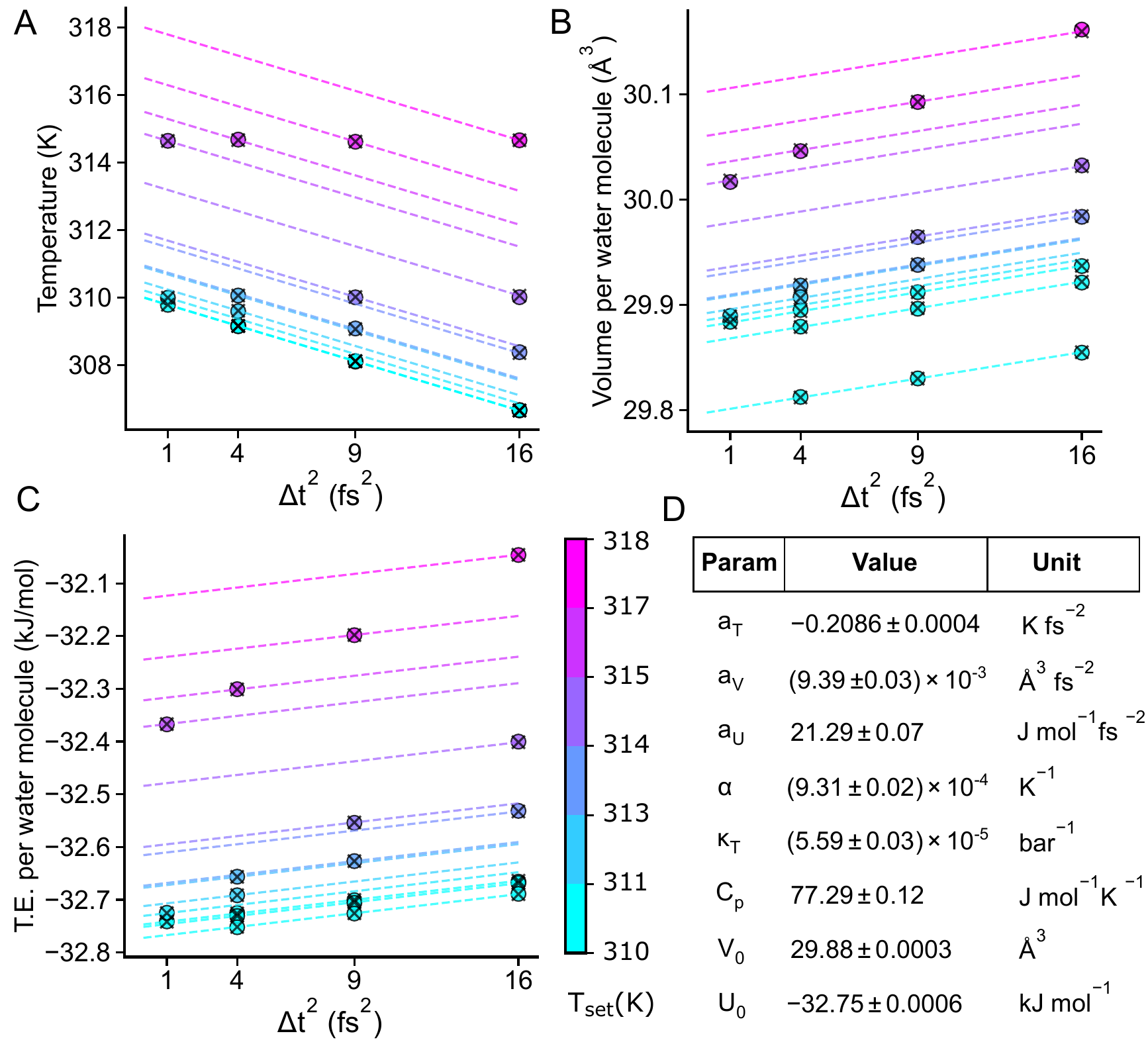}
\vspace{-4mm}
\caption{
\textbf{Time step dependence of temperature, volume, and energy.}
(A) Simulation averaged temperature, (B) volume per water molecule, and (C) total energy (T.E.) per water molecule, respectively, as functions $\Delta t^2$. 
Circles represent time-averaged values from individual simulations performed at different combinations of \{$\Delta t$, $T_\mathrm{set}$, $P_\mathrm{set}$\}.
Symbol color represents $T_\mathrm{set}$, defined in the color bar.
The cross symbols ($\times$) indicate corresponding predictions from the fitted thermodynamic model evaluated at the simulated conditions.
Dashed lines represent the continuous model predictions for each unique combination of \{$T_\mathrm{set}$, $P_\mathrm{set}$\} across the full range of $\Delta t^2$ values.
(D) Fitted model parameters calculated per water. Uncertainties correspond to standard errors derived from the covariance matrix of the weighted global least-squares fit.
}
\label{fig:fig2}
\end{figure}
Time-averaged temperature ($T$), volume ($V$), and total energy ($U$) obtained from independent simulations performed at different combinations of \{$\Delta t,~T_\mathrm{set},~p_\mathrm{set}$\} were fit to the thermodynamic model, Eqs.~\eqref{eq:T}, \eqref{eq:U}, and \eqref{eq:V}. 
Global optimization of the eight model parameters in the model \{$U_0$, $\kappa_T$, $V_0$, $C_p$, $\alpha$, $a_U$, $a_V$, $a_T$\} yields predictions in excellent agreement with the simulated data across a wide range of set temperature and pressure~(\fig{fig2}A-C).
Interestingly, the global fit yields experimental observables $\alpha$, $\kappa_T$ and $C_p$~(\fig{fig2}D) in agreement with reported values for the standard-TIP3P (sTIP3P) water model,\cite{IZAD2014} of $9.2~\mathrm{K^{-1}}$, $5.74\times 10^{-5}~\mathrm{bar^{-1}}$ and $78.4~\mathrm{J~mol^{-1}~K^{-1}}$, respectively. 
Minor differences ($<~2.5~\%$) likely arise from the use of CHARMM-modified TIP3P water instead of sTIP3P model. 
The thermodynamic model thus successfully captures the systematic drift in thermodynamic variables arising from finite time step integration, with error terms up to $\mathcal{O}(\Delta t^2)$ being sufficient. 
Once the model parameters are determined for a given system, observables can be reliably extrapolated to the zero time-step limit, which cannot be achieved through direct numerical integration.
Consequently, the model mitigates discretization bias and enhances the accuracy of observables calculated from MD simulations.

When using multi-time stepping (MTS), we noticed that the time-averaged values of observables ($U, V, T$) remain close to those obtained without MTS; therefore, the same model can be used, taking $\Delta t$ to be the inner (fast) time step used to evaluate the rapidly varying short-range interactions. 
Additional MTS data points were excluded from the model fit to avoid over-weighting near-duplicate conditions.
Simulation input parameters for the additional runs, with corresponding time-averaged observables are reported in SI Table 3.

\subsection{Ensemble correction for time step dependence and deviations in thermodynamic state}

\begin{figure*}[t]
\centering
\includegraphics[width=1\textwidth]{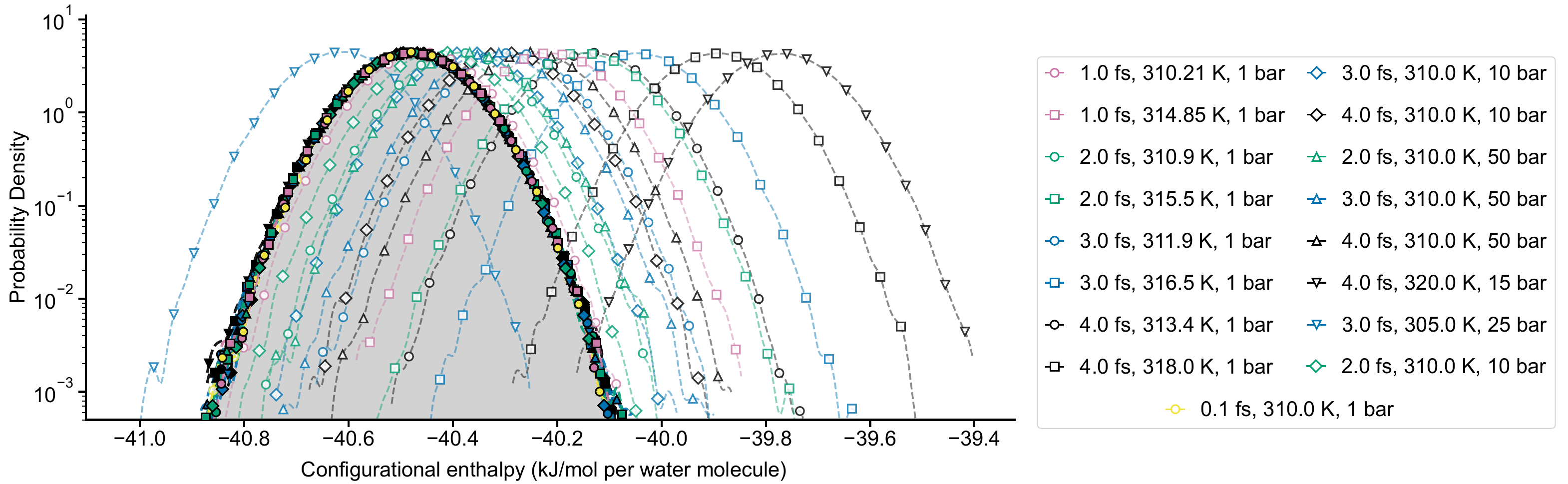}
\vspace{-4mm}
\caption{
\textbf{Thermodynamic model collapses disparate distributions of enthalpy.}
Probability density of the configurational enthalpy per water molecule, for the system of $N=5184$ TIP3P water molecules, plotted on a semi-logarithmic scale. 
Faint dashed lines with open markers show the raw distributions from simulations performed at various time steps ($\Delta t = 0.1$--$4.0$ fs), temperatures ($T_\mathrm{set} = 310$--$318$ K), and pressures ($p_\mathrm{set} = 1$--$50$ bar).
Solid markers with black outlines represent the same data after correction to a common reference state ($\Delta t \to 0$, $T_{\mathrm{ref}}=310$ K, $p_{\mathrm{ref}}=1$ bar) using the fitted thermodynamic model. 
The gray shaded area represents a Gaussian distribution centered at the theoretical zero time step mean enthalpy, $H_0 = U_0 + p V_0$, with a variance determined by the fitted heat capacity, $\sigma_{H_\mathrm{conf}}^2 = k_B T_{\mathrm{ref}}^2 C_{p_\mathrm{conf}} / N$, where $C_{p_\mathrm{conf}} = C_p -3R$.
}
\label{fig:fig3}
\end{figure*}

The fitted thermodynamic model effectively corrects the systematic drift in the mean observables, enabling extrapolation to the limit of zero time step. Its robustness is further validated by applying the model-predicted mean corrections ($\Delta U, \Delta V$) to shift observable distributions from ($\Delta t_{\text{sim}}, T_{\text{sim}}, p_{\text{sim}}$) to ($\Delta t_{\text{ref}}=0, T_{\text{ref}}, p_{\text{ref}}$).
This test places a strict demand on the framework: since the model parameters were derived solely from time-averaged quantities (first moments), reliable overlap requires that the predicted energy differences correctly map the full probability distribution of one state onto another.

The corrected distributions of configurational enthalpy obtained from simulations across a wide range of time steps, temperatures, and pressures all exhibit excellent overlap (Kullback–Leibler divergence below $10^{-3}$) when mapped to an otherwise arbitrary target reference state, at $T=310~$K, $p=1~$bar, and $\Delta t \to 0$~(\fig{fig3}).
Although the time-step-dependent error is already small at 1 fs, we simulated an additional system at $\Delta t$~=~0.1~fs for 100~ns, to pinpoint the target distribution more accurately~(\fig{fig3}: yellow points).
Strikingly, the corrected distributions agree perfectly even where the raw sampled distributions share negligible overlap.
This success highlights a key feature of the system: within the range of parameters explored, the second moments of the distributions--which are related to thermodynamic quantities such as heat capacity and compressibility--remain largely consistent. 
Consequently, the correction amounts to a linear shift of the respective distribution by the model-predicted mean offsets (e.g., $\Delta U$ and $\Delta V$). The ability to bridge disparate simulation states indicates that the thermodynamic model captures the macroscopic energetic shifts governing the system, validating its predictive power beyond mean-value extrapolation. 
Thus, we can generate effective Boltzmann samples at the target thermodynamic state for any observable explicitly represented in the thermodynamic model by translating the corresponding source-trajectory time series with the model-predicted mean corrections. 

\subsection{Fitting the model to protein simulations}

\begin{figure}[thbp!]
\centering
\includegraphics[width=0.46\textwidth]{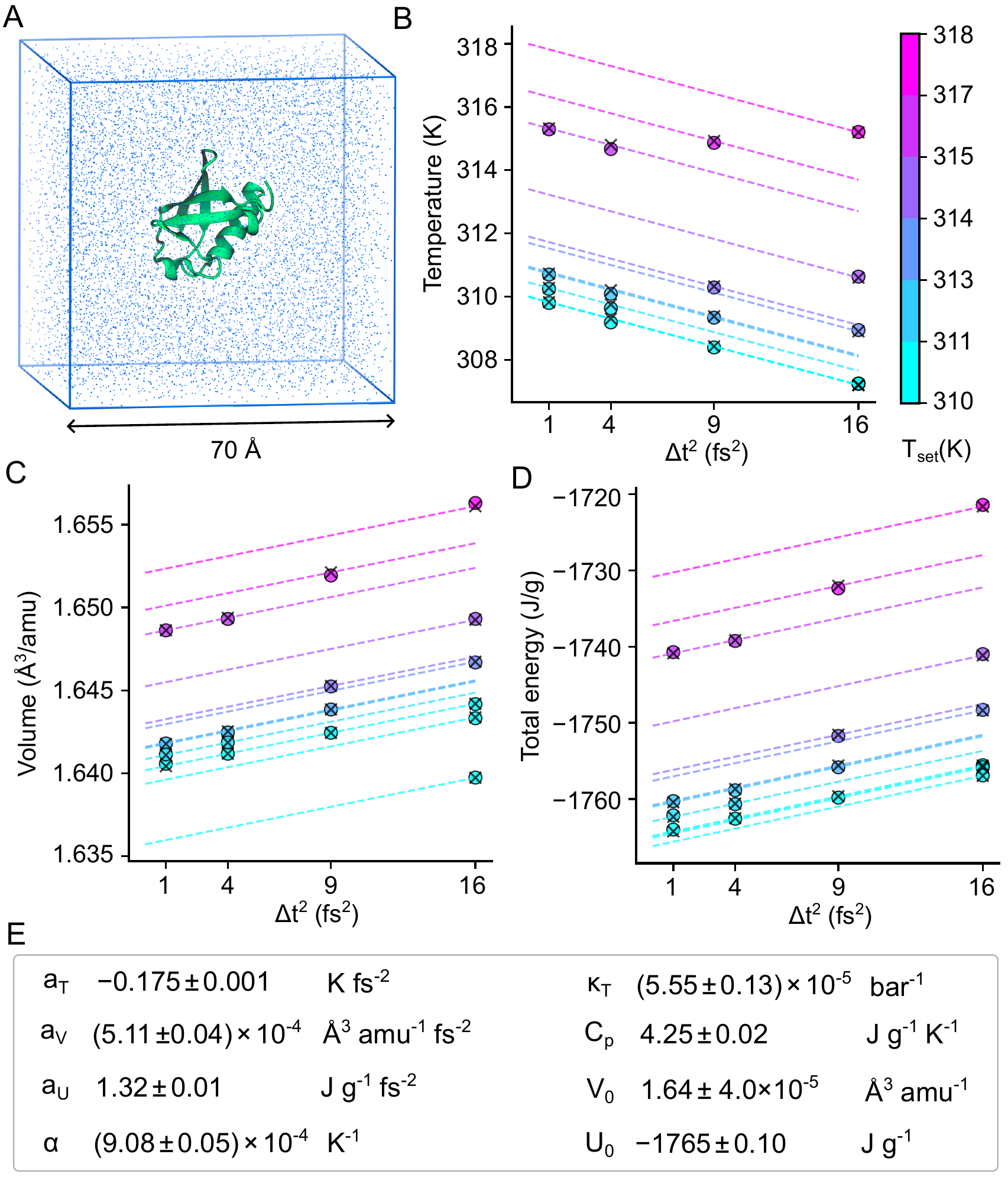}
\caption{
\textbf{Time step dependence of the thermodynamic model for a protein in solution.}
(A) Ubiquitin molecule (PDB ID: 1UBQ) in a water box.
(B) Simulation averaged temperature, (C) volume per unit mass ({\AA}$^3$/amu), and (D) total energy (T.E.) per unit mass (J/g), respectively, as functions of $\Delta t^2$. 
Symbols, colors, and dashed lines carry same meaning as in~\fig{fig2}A-C.
(E) Fitted model parameters. Uncertainties correspond to standard errors derived from the covariance matrix of the weighted global least-squares fit.
}
\label{fig:fig4}
\end{figure}

To test the applicability of the thermodynamic model for biomolecular simulations, we fit it to a set of simulations of a ubiquitin protein placed in aqueous solution~(\fig{fig4}A).
As expected, the thermostat yields a simulation temperature consistently lower than $T_\mathrm{set}$, with the magnitude of the drift depending on the time step of the simulation~(\fig{fig4}B).
This discretization error results in a systematic shift also in other thermodynamic quantities, including the system volume and energies~(\fig{fig4}C). The thermodynamic model fits perfectly to the collection of simulations exhibiting that the model fits perfectly and that the zero time step estimates can hence be reliably extrapolated. Interestingly, the temperature coupling, given by $a_T$, is weaker than the case with pure water simulations. 

For the case of the simple harmonic oscillator with frequency $\omega$, we know that with a second-order symplectic integrator the leading discretization error in thermodynamic observables will scale as $(\omega \Delta t)^2$, where $\Delta t$ is the time step of integration.\cite{PAST1988}
Generally speaking, the magnitude of the temperature drift will be determined by the frequency spectrum of the unconstrained degrees of freedom in the system, albeit with higher-frequency modes contributing more to the error.
In our simulations involving rigid water, the
fastest unconstrained degrees of freedom in the pure water system are the rotational librations (hindered rotations), which form a broad band between approximately 400 and 900 cm$^{-1}$ (12–27 THz).\cite{CARE1998}
By contrast, the solvated protein introduces a dense spectrum of low-frequency modes ($<$ 200 cm$^{-1}$) associated with collective domain motions and backbone torsions. While high-frequency bond vibrations exist within the protein (e.g., C-C, C-O etc.), they are sparse compared to the high density of fast water librations, displaced by the macromolecule. 
Since the discretization-induced temperature shift is effectively an average over the entire frequency spectrum weighted by $\langle \omega^2 \rangle$, the pure water system at the same total mass should exhibit a larger global temperature deviation than the protein-water system.

Interestingly, the sensitivity of the system volume to the time step, $a_V$~(\fig{fig4}C), is consistent between the two systems once differences in units are accounted for. 
For pure water, we calculated the specific volume per molecule, yielding $a_V$ in units of {\AA}$^3$ fs$^{-2}$~(\fig{fig2}D), whereas, for the heterogeneous protein system we calculate the specific volume per unit mass, yielding $a_V$ in units of \AA$^3$ amu$^{-1}$ fs$^{-2}$~(\fig{fig4}E). 
Multiplying by 18.05 amu (mass of the dominant system constituent by mass, i.e., water), the resulting $a_V$ is close to that fitted for pure water. 
Similarly, other quantities such as $\alpha$, $\kappa_T$, and $C_p$ are also remarkably similar. 
We therefore find that the thermodynamic model is robust also across systems and can be used to globally fit biomolecular systems and correct for time step errors.

\subsection{Validation of ensemble correction in the protein system.}

\begin{figure*}[thbp!]
\centering
\includegraphics[width=1\textwidth]{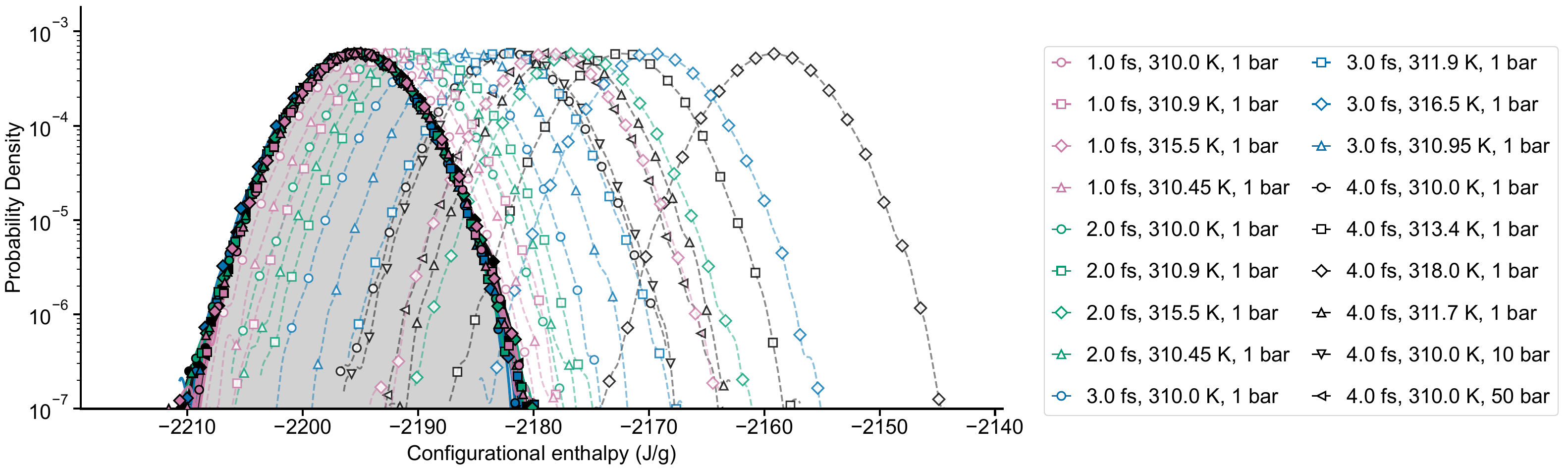}
\vspace{-4mm}
\caption{
\textbf{Thermodynamic model collapses disparate distributions of enthalpy for protein in water.}
Probability density distributions of the configurational enthalpy per unit mass for the protein system, plotted on a semi-logarithmic scale. 
Symbols, lines, and the shaded region carry same meaning as defined in~\fig{fig3}. Marker shape and color distinguish different systems, as per legend. 
}
\label{fig:fig5}
\end{figure*}

Analogous to the pure water analysis~(\fig{fig3}), we used the fitted parameters from the thermodynamic model~(\fig{fig4}E) to correct the distributions of configurational enthalpy obtained from simulations across a wide range of time steps, temperatures, and pressures.
The resulting distributions exhibit excellent overlap when corrected to the target reference state at $T=310~$K, $p=1~$bar, and $\Delta t \to 0$~(\fig{fig5}).
Thus, we find that this correction strategy is effective also for biomolecular systems, allowing one to recover effective Boltzmann samples of target thermodynamic properties just as in the case of pure water.

\section{Conclusions}

Physical observables obtained from molecular dynamics simulations noticeably depend on the integration time step, with the resulting systematic errors scaling as $\mathcal{O}(\Delta t^2)$ for the widely used Verlet-type integration schemes. This dependence is particularly relevant given the current demand for high-throughput simulations using aggressive time steps, e.g., 4 fs with hydrogen mass repartitioning. 
Our results highlight that widely used integrators, such as the BBK scheme implemented in popular codes like NAMD and AMBER, can suffer from significant temperature deviations, where the simulated system temperature drops appreciably below the thermostat setpoint as the time step increases.
These artifacts arise because the discretized propagator does not sample the exact Boltzmann distribution of the physical Hamiltonian, but rather a distribution defined by a conserved shadow Hamiltonian.

While directly correcting the shadow Hamiltonian is non-trivial,\cite{Skeel:2001} we show that the resulting deviations in observables follow predictable thermodynamic laws. By fitting a simple linear thermodynamic model to data from simulations with finite time steps, we can robustly extrapolate quantities of interest (such as energy, volume, and temperature) to the ideal zero time step limit, effectively removing the discretization bias.

A seeming flaw in the numerical integration of the equation thus gives us access to thermodynamic information. The framework can then be used to extrapolate to zero time step and to thermodynamic states of interest. In particular, we can use it to generate effective Boltzmann samples from biased finite time step trajectories. By applying correction factors based on the thermodynamic model, one can recover target probability distributions simply by shifting the observed sample distributions. This capability is expected to be particularly valuable for enhanced sampling techniques like replica exchange MD\cite{Okamoto:CPL:1999:remd} and free energy methods such as umbrella sampling,\cite{Torrie:74} where the assumption of exact Boltzmann sampling at a particular temperature is critical for the validity of thermodynamic reweighting and the convergence of free energy estimates.\cite{Rosta:JCTC:2009} Systematic deviations from Boltzmann sampling have been found to break detailed balance in case of replica exchange and to alter protein folding thermodynamics in an uncontrolled manner.
As we showed here for water and a protein in water, the extrapolation to zero time step allows us to consistently map simulations for a wide range of time steps and from a range of thermodynamic states onto a defined thermodynamic state.

\vspace{1mm}

\begin{acknowledgments}
The authors thank the Max Planck Society for support and the Max Planck Computing and Data Facility for computational resources.
K.C. thanks Christopher Maffeo and J\"urgen K\"ofinger for valuable discussions. G.H. thanks Christoph Dellago for many discussions, inspiration, and sharing his deep insights and understanding. 
\end{acknowledgments}

\vspace{0.5cm}
\textit{Conflicts of interest - } None declared.

\section*{References}
\bibliographystyle{apsrev4-2}
\bibliography{bib_1}

\end{document}